\documentclass[12pt]{article}
\usepackage{amsmath}
\usepackage{amssymb}
\usepackage{amsthm}
\usepackage{graphicx}
\usepackage{psfrag}
\usepackage[hang, nooneline]{subfigure}
\usepackage{color}

\usepackage{soul} %added for \hl
\usepackage{cite}

\usepackage{amssymb} 
\usepackage{amsfonts} 
\usepackage{amsmath} 
\usepackage{slashed}

\usepackage{egothic}%
\usepackage{pgothic}%
\usepackage{yfonts}%

 \usepackage{yhmath} %%the simplest solution for wider \widehat{}!!!!!

\newcommand{\we}{\wedge}

\newcommand{\der}{\partial}

\newcommand{\inn}{\hspace*{2pt}\raisebox{-1pt}{\rule{6pt}{.3pt}\hspace*
{0pt}\rule{.3pt}{8pt}\hspace*{3pt}}}

\newcommand{\beq}{\begin{equation}}
\newcommand{\eeq}{\end{equation}}
\newcommand{\beqa}{\begin{eqnarray}}
\newcommand{\eeqa}{\end{eqnarray}}
\newcommand{\nn}{\nonumber}

\newcommand{\pbr}[2]{ \{ \hspace*{-2.6pt} [ #1 , #2\hspace*{1.4 pt} ] 
\hspace*{-2.6pt} \} }
\renewcommand{\ge}{\textgoth{e}}

\author{ 
\vspace*{1ex} \\
Monika E. Pietrzyk$^{1}$   and C\'ecile Barbachoux$^{2}$ 
\\ \small $^{1}$ Mathematics and Physical Sciences, University of Exeter, 
 EX4 4QL Exeter, UK
\\ 
\small {\!\!\!}${\!\!\!}^{2}$  
Sciences and Technologies Department, C\^ote d'Azur University/INSPE,
06000 Nice, France 
}

\date{} 

\title{On the  Covariant Hamilton-Jacobi Equation for the Teleparallel Equivalent of General Relativity }

\begin{document}
\maketitle
\begin{abstract}
The covariant Hamilton-Jacobi equation for the Teleparallel Equivalent of General Relativity 
 is derived based on the analysis of constraints within the De Donder-Weyl covariant 
 Hamiltonian theory developed by Kanatchikov. 
\end{abstract}

\section{{ Introduction}} 
The covariant Hamilton-Jacobi equation for General Relativity in metric variables has been found by 
Th\'eophile De Donder in 1930 \cite{dedonder}  and rediscovered by Petr Ho\v{r}ava in 1991 \cite{horava}.  In 1935 Hermann Weyl has considered a Hamilton-Jacobi theory for multidimensional variational problems \cite{weyl}  based on earlier ideas of Volterra \cite{volterra}, Carath\'eodory \cite{carath} 
and the  G\"ottingen school of Hilbert \cite{prange}. A good historical review 
with many references can be found in the book by Hanno Rund \cite{rund}.  
In the 1980-es the interest to such theories has been revived by Hans Kastrup in his comprehensive review paper \cite{kastrup} and Yoichiro Nambu \cite{nambu} in the context of classical string dynamics. The connection between the geodesic fields \cite{boerner} in the 
covariant Nambu-Hamilton-Jacobi theory of strings and classical fields described by simple bivectors has been discussed in \cite{kastruprinke,rinke,ikan-string}. 
 %A few examples of applications in field theory have been worked out in \cite[vonrieth]
The geometric formulations using the language of differential geometry have been developed later e.g. in  \cite{dlhj1,dlhj2,dlhj3,rund89,bruno-hj,rr21}. The relation to the standard Hamilton-Jacobi theory in field theory which is using an explicitly distinguished time dimension and canonical Hamiltonian formalism 
\cite{weiss,hj-thesis} has been discussed only in three papers \cite{ik-pla}, \cite{nikolic} and \cite{riahi} and it still remains a subject worthy of further investigation. 
Let us note that, whereas the covariant Hamilton-Jacobi equations in modern field theory and general relativity are relying on an additional structure of a global foliation with space-like leaves \cite{rovelli,dopli,conrady}, the covariant Hamilton-Jacobi theories in the calculus of variations 
do not introduce any additional structure on the space of independent space-time variables and treat them  equally.    

The covariant Hamilton-Jacobi theories of multidimensional variational problems which describe classical fields have naturally lead to the question of whether a formulations of quantum theory for fields 
may naturally reproduce the corresponding Hamilton-Jacobi theories in the quasiclassical limit. 
The modern answer to this question is given 
by the programme of precanonical quantization which started in \cite{ik1,ik2,ik3,ik4,ik5,ik5e} using the 
earlier discovery of the bi-graded (Gerstenhaber) analogue of the Poisson bracket in the 
simplest De Donder-Weyl (DW) covariant Hamiltonian formalism \cite{ikbr1,ikbr2,ikbr3,khbr1,khbr2,khbr3}.
 This formalism is related to the covariant Hamilton-Jacobi theory discussed by Th. De Donder \cite{dedonder} and H. Weyl \cite{weyl} in the same way as the standard Hamilton-Jacobi theory in mechanics is related to the Hamiltonian formalism \cite{rr21}. Later on precanonical quantization has been applied to gauge fields \cite{ik5e,iky1,iky2,iky3} and general relativity in the standard metric formulation  \cite{ikm1,ikm2,ikm3,ikm4} 
 and the Palatini vielbein formulation \cite{ikv1,ikv2,ikv3,ikv4,ikv5}. 
The relation of this approach with the standard canonical quantization in the Schr\"odinger representation has been established in \cite{ik-pla,iks1,iks2,iksc1,iksc2,iksc3,iky1,iky2} 
  both for quantum scalar fields in flat and curved space-times and Yang-Mills gauge theory. 

The classical limit of precanonical quantization has been analysed by means of a generalization of the Ehrenfest theorem \cite{ik3,ik4,ik5e,ikv4}: the  De Donder-Weyl covariant Hamiltonian field equations 
are shown to be the equations satisfied by the expectation values of certain operators. 
However, the reduction of precanonical Schr\"odinger equation to the De Donder-Weyl covariant Hamilton-Jacobi equation so far has been demonstrated only for scalar fields in \cite{ik3,guiding}. 

In this paper we would like to extend the earlier work by De Donder\cite{dedonder} 
and Ho\v{r}ava\cite{horava} to the case of the teleparallel equivalent of general relativity (TEGR). The interest in teleparallel theories of gravity has been revived recently due to their theoretically attractive features such as a close connection with the Poincar\'e gauge theory of gravity, a possibility of defining the energy-momentum tensor of the field of gravity, and potential applications in the field of cosmology and gravitational waves. It has stimulated numerous generalizations and theoretical developments in different directions. Among those the attempts to analyze the canonical Hamiltonian formalism and the complicated structure of constraints have been made which are a prerequisite for canonical quantization. 
To our knowledge, no Hamilton-Jacobi theory based on the canonical Hamiltonian analysis has been 
formulated for teleparallel gravities so far despite it could be a bridge or a hint to a quantum formulation, as it happened with the canonical Hamilton-Jacobi formulation of General Relativity by 
Asher Peres in 1962 \cite{peres2} serving as an inspiration for the Wheeler-De Witt equation in canonical quantum gravity. As the precanonical quantization of TEGR has been recently presented in 
\cite{ik-tpq,ik-mg21}\footnote{We thank the author of these conference talks for sharing his slides prior to the publication of the Proceedings.}, 
we are interested to formulate its covariant Hamilton-Jacobi counterpart as a potential testing ground 
of the ability of this quantum formulation of teleparallel gravity to reproduce the correct classical limit %%of the precanonical quantization . 
in the quasiclassical approximation in a similar way to the derivation of the 
classical equations of general relativity from the quantum geometrodynamics by Ulrich Gerlach \cite{gerlach69}. 
Here we will limit our consideration to the case of the TEGR and leave a consideration of more general 
teleparallel theories of gravity to a later occasion.

\bigskip

\section{Palatini formulation of TEGR} 

In spite of many different formulations of TEGR in metric and vielbein variables \cite{maluf2,jhk} 
it is the Palatini formulation by Maluf\cite{maluf}  which has been successfully applied for the 
covariant Hamiltonian analysis of TEGR in the sense of De Donder-Weyl formulation \cite{ik-tpq,ik-mg21}. 
In this formulation the vielbeins $e^\alpha_a$ and the variables $t_{abc}= - t_{acb}$ are the 
independent field variables. Using the Lagrangian density in the form 
\beq  \label{lagrtp}
L = \frac{1}{16\pi G} \ge \Sigma (t)^{abc} \left(t_{abc}-2 T_{abc}\right)
\eeq
where  
\begin{align}
T^{c}{}_{ \alpha\beta} &:= \der_\alpha e^c_\beta - \der_\beta e^c_\alpha \nn \\
T^c{}_{ab} &:= e^\alpha_a e^\beta_b T^c{}_{\alpha\beta}, \quad T_b := T^a{}_{ab}  \\ 
\Sigma (t)_{abc}&:=  \frac12 (\eta_{ac} t_{b} - \eta_{bc} t_{a})+   \frac14 (t_{abc}+t_{bac}-t_{cab})   \nn 
\end{align}
and $\ge := \det (e^a_\alpha)$. The variation of  $t_{abc}$ identifies the field $t_{abc}$ on classical solutions as the torsion field $T_{abc}$. The variation of vielbeins 
$e^\alpha_a$ reproduces the Einstein equations in vielbein variables \cite{maluf}.

%The Euler-Lagrange field equations yield 
%\beqa
%\delta t:&& \quad \Sigma (t)^{abc} = \Sigma (T)^{abc}  \Rightarrow t_{abc} = T_{abc} 
%\label{deltaf}\\
%\delta e:&& \quad Einstein  \; equations
%\label{deltae}
%\eeqa

 \section{De Donder-Weyl covariant Hamiltonian analysis of TEGR}  
 
 \subsection{De Donder-Weyl Hamiltonian formulation in flat space-time}
 
A  Lagrangian function $L(y, y_\alpha, x^\alpha)$ can be viewed as a 
a function of space-time variables,  $x^\alpha$, 
%or independent variables in the language of the calculus of variations,   
the field variable, $y$, and the first-jet coordinates, $y_\alpha$. 
On classical field configurations $y = y(x)$ and  $y_\alpha = \der_\alpha y(x)$, where 
$\der_\alpha$ is a short-hand notation for $\frac{\der}{\der x^\alpha}$. 
The variation of the field configuration $y(x)$ yields the Euler-Lagrange field equations. 
The De Donder-Weyl Hamiltonian formulation of the latter is based on a new set of variables 
\beq 
p^\alpha_y := \frac{\der L}{\der y_\alpha} 
\eeq
called polymomenta corresponding to the field variable $y$
and 
\beq
H := p^\alpha y_\alpha - L =: H\left(y, p_y^\alpha, x^\alpha \right)
\eeq 
called the De Donder-Weyl or DW Hamiltonian density. In terms of new variables the 
Euler-Lagrange field equations are cast in the first order Hamiltonian form 
\begin{align}
\der_\alpha y &= \frac{\der H}{\der p_y^\alpha} \label{dw1}\\
\der_\alpha p_y^\alpha &=  - \frac{\der H}{\der y} \label{dw2}
\end{align}
This transformation is possible when the Lagrangian density is regular in the sense that the determinant of the matrix $W^{\alpha\beta} = \frac{\der^2 L}{\der y_\alpha \der y_\beta}$ is nonvanishing. 

The Hamilton-Jacobi equation corresponding to the Hamiltonian equations (\ref{dw1}), (\ref{dw2}) 
can be obtained in the form of the first order partial differential equation for eikonal functions 
on the covariant field configuration space $S^\alpha (y,x)$: 
\beq
\frac{\der S^\alpha}{\der x^\alpha} + H \left(y, \frac{\der S^\alpha}{\der y}, x^\alpha \right) = 0 
\eeq 
which describes the solutions of the Euler-Lagrange equation $y = y(x)$ as geodesic fields in the 
 space of field variable $y$ and space-time variables $x^\alpha$ which are 
gived by the equations called the embedding conditions 
\beq 
p^\alpha_y (y(x), y_\alpha(x)) = \frac{\der S^\alpha}{\der y} \label{dwhjeq}
\eeq
Here the left hand side is the explicit expression of polymomenta $p^\alpha$ 
in terms of Lagrangian variables $(y, y_\alpha)$ taken along a classical configuration $y(x)$. 

\subsection{The constraints}

Let us apply the procedure of the covariant DW Hamiltonian theory to the Lagrangian density (\ref{lagrtp}). From the expressions of the polymomenta corresponding to the independent field variables $e$ and $t$ we immediately see that they are constrained. Namely, 
\beqa
%& C^\mu_{t_{abc}} & := 
p^\mu_{t_{abc}} &\approx 0, \label{cf}\\
%& C^\mu_{e^a_\beta}\; & :=  
p^\mu_{e^a_\beta} +  \frac{1}{4\pi G}\ge \Sigma(t)_a{}^{bc} e^{[\mu}_b e^{\beta]}_c &\approx 0 
\label{ce}
\eeqa
The first set of constraints will be denoted $C^\mu_{t_{abc}}$ and the second one  $C^\mu_{e^a_\beta}$, 
the symbol $\approx$ stands for "weakly equals to" according to the tradition set by P. Dirac in his 
theory of constraints \cite{dirac}.  The appearance of constraints is problematic for the straightforward 
application of DW Hamiltonian formalism. However, Kanatchikov \cite{ik-dirac} has generalised the Dirac's constraints algorithm to the covariant DW Hamiltonian theory which was applied to the Palatini formulation of vielbein gravity \cite{ikv1,ikv2,ikv3,ikv4,ikv5} and to other modified theories of gravity 
\cite{mx1,mx2,mx3}.  

In the follow-up we will often use a condensed notation 
that the indices $e$ and $t$ mean the whole set of indices carried by these fields and their repetition 
means summation over all those indices according to the Einstein summation rule.   In this notation 
the primary constraints read 
\beq 
C^\alpha_t \approx 0, \; C^\alpha_e \approx 0 
\eeq
and the primary DW Hamiltonian density is obtained in the form 
\beq \label{dwhprim}
{\ge H} := p^\mu_t \der_\mu t + p^\mu_e \der_\mu e - L 
 %%\approx  - \frac{1}{16\pi G} \ge \Sigma (t)^{abc} t_{abc} 
\approx \frac14 p^\mu_{e^a_\beta} e_\mu^b e_\beta^c t^a{}_{bc}  
\eeq
Kanatchikov\cite{ik-dirac} has developed an algorihtm of the constraints analysis in the covariant 
DW Hamiltonian formalism. Its central idea is to use the forms of constraints and their brackets. 
The forms of constraints are given by 
\beq 
C_e:=  C^\mu_{e}\upsilon_\mu,  \quad C_t:=  C^\mu_{t}\upsilon_\mu 
\eeq
where $\upsilon_\mu := \der_\mu \inn\! (dx^1\we...\we dx^n)$ if the space-time is $n$-dimensional. 
The brackets of forms are defined by the polysymplectic form on the 
unconstrained phase space of polymomentum and field variables 
\beq
\Omega =  d e \we dp^\mu_e \we\upsilon_\mu + d t \we d p^\mu_t \we \upsilon_\mu
\eeq
$\Omega$ maps a form $C$ of degree $(n-1)$ to a vector field $\chi_C$
\beq
\chi_C \inn \Omega = d C
\eeq 
and the bracket of two $(n-1)$-forms is defined as 
\beq 
\pbr{C_1}{C_2} := \chi_{C1}\inn d C_2. 
\eeq 
Note that this definition produces brackets for a limited class of forms called Hamiltonian forms.  
 The Lie algebra structure defined by this bracket is embedded into a bigger bi-graded structure defined 
on forms of all degrees below $(n-1)$ which is known as the Gerstenhaber algebra 
\cite{ikbr1,ikbr2,ikbr3,khbr1,khbr2,khbr3}. 

A straigrtforward calculation yields the following brackets of forms of constraints: 
\beqa
&&\pbr{C_{e^a_\alpha}}{C_{e^d_\beta}} =: C_{e^a_\alpha e^d_{\beta}} = \frac{1}{4\pi G} \der_{e^a_\nu}\left( \ge \Sigma(t)_d{}^{bc} e^{[\alpha}_b e^{\beta]}_c \right)
\upsilon_\nu 
 %%\quad\mathrm{or}\quad C_{e e'} 
\label{cece} \\
&&\pbr{C_{e^a_\beta}}{C_{t_{dbc}}}=: C_{e^a_\beta t_{dbc}} = -\frac{1}{4\pi G}\ge e^{[\nu}_k e^{\beta]}_l 
\frac{\der\Sigma(t)_a{}^{kl}}{\der {t_{dbc}}} \upsilon_\nu
  %%\quad\mathrm{or}\quad C_{e t}  
\\
&&\pbr{C_t}{C_{t'}}=: C_{tt'} = 0  
\eeqa
%Note that $ C_{e e'} $ depends on $e$ and $t$ whereas $ C_{e t} $ depends only on $e$. 
It shows that we have a second-class constraints according to Dirac's classification \cite{dirac}
a proper generalization of Poisson bracket is needed to account for such constraints.

\subsection{Generalized Dirac brackets}

A generalization of Dirac bracket in the context of constrained DW Hamiltonian systems with 2-nd class constraint has been proposed by Kanatchikov in \cite{ik-dirac}. It has been  
used in his precanonical quantization of Einstein gravity in vielbein variables \cite{ikv1,ikv2,ikv3,ikv4,ikv5}.
For  $(n-1)$-- and $0$--  forms $A$ and $B$ 
\beq \label{pdir}
\pbr{A}{B}{}^D := \pbr{A}{B}{} - \Sigma_{U,V}\pbr{A}{C_U}{}\bullet  C_{UV}^{\sim 1} \we \pbr{C_V}{B}{} 
\eeq
The indices $U,V$ enumerate the primary constraints, 
\beq
A\bullet B := *^{-1}(*A\we*B) 
\eeq 
and the pseudoinverse matrix $C^{\sim 1} =C^{\sim 1}_\mu dx^\mu $ is defined by the relation 
\beq
C^{\sim 1}\bullet C^{} \we C^{\sim 1} = C^{\sim 1} 
\eeq
where the distributive law for $\we$ and $\bullet$ products is that the wedge product $\we$ acts first. 
The result of the calculation of generalized Dirac brackets\cite{ik-mg21} reads 
\begin{align}
\pbr{p_t}{t'}{}^D &= 0 \label{ptt}
\\ 
\pbr{p_t}{p_{t'}}{}^D &= 0 %\pbr{p_t}{p_{t'}}  - \pbr{p_t}{ C_e} C^{\sim 1}_{e t''} \pbr{C_{{t''}}}{t'} = 
\\
\pbr{p_t}{e}{}^D &= 0 %\pbr{p_t}{t'}{} - \pbr{p_t}{ C_e} C^{\sim 1}_{e t''} \pbr{C_{{t''}}}{t'} = 
\\
\pbr{p_t}{p_{e}}{}^D &= 0
\\
\pbr{p_e}{p_{e'}}{}^D &= 0 \label{pepe}
\\
\pbr{e \upsilon_\mu}{e'_{}}^D &
%=  C^{\sim 1}_{\mu ee'} 
= 0 \label{eve}
\\
\pbr{p_e}{e'}{}^D &= 
 %\pbr{p_e}{e'}{} -  \pbr{p_e}{C_{e''}} C^{\sim 1}_{e'' t''} \pbr{C_{t''}}{e'}  = 
 %\pbr{p_e}{e'}{}=
 \delta_{ee'}
\label{pee}
\\
%\pbr{p^\mu_e}{e'\upsilon_\beta}{}^D 
%%&= \pbr{p^\mu_e}{e'\upsilon_\beta}{} -  \pbr{p^\mu_e}{C_{e''}} C^{\sim 1}_{e'' t''} \pbr{C_{t''}}%{e'\upsilon_\beta}  
%%\nn \\ 
%& =  \pbr{p^\mu_e}{e'\upsilon_\beta}{}=\delta_{ee'}\delta^\mu_\beta
%\label{pmuee}
%\\
\pbr{t}{p_e}{}^D &=
%- \pbr{p_e}{t}{} +  \pbr{p_e}{C_{e'}} \bullet ( C^{\sim 1}_{e' t'} \we \pbr{C_{t'}}{t} ) 
%\nn \\ & \quad 
%=  \frac{1}{4\pi G} \der_e \left(\ge{} \Sigma(t)_b{}^{cd} e{}_c^\alpha e_d^\beta \right)  
%C^{\sim 1}_{\alpha e^b_\beta t} = 
 C^\alpha_{e e'} C^{\sim 1}_{\alpha e' t}
\label{tpe}
\\
 \pbr{t \upsilon_\alpha}{t'}^D &
 = C^{\sim 1}_{\alpha t t'} 
 \label{tt}
 \\
\pbr{e \upsilon_\alpha}{t_{}}^D &
= C^{\sim 1}_{\alpha et}
\label{et}
\end{align} 
From (\ref{ptt}) we conclude that variables $t$ have vanishing Dirac brackets with their conjugate polymomenta and from (\ref{ptt}) we conclude that the Dirac brackets between different components of $t$ are not vanishing. 
It means that the unconstrained polymomentum phase space of variables $(e,t,p^\alpha_e,p^\alpha_t)$ 
is reduced to the space of vielbeins $e$ and their polymomenta $p_e$ where the 
variables $t$ are functions on this reduced polymomentum phase space $t = t(e,p_e)$ such as 
their Dirac brackets with the variables of the reduced polymomentum phase space are given by 
 (\ref{tpe}) and (\ref{et}).

\subsection{Functions $t_{} (e,p_e)$ and $H(e,p_e)$ on the reduced polymomentum phase space} 

The analysys of constraints in the previous section has shown that the primary polymomentum phase space 
of variables $(e,t,p_e,p_t)$ is effectively reduced to the space of variables $(e,p_e)$ and 
the auxiliary fields $t$ which we have introduced in the Lagrangian Palatini formulation (\ref{lagrtp}) 
are becoming functions of the reduced polymomentum phase space. Then the DW Hamiltonian density  
also becomes a function on the reduced polymomentum phase space. Here we present the explicit expressions 
of those functions. 

%Let us express $t_{abc}$ as a function of $e$ and $p_e$. From (\ref{ce}) 
%it follows that it is possible if the map $t \rightarrow \Sigma$ is invertible. 
From (2) we obtain 
\beq
t_{abc} + \eta_{ac} t_b -\eta_{ab} t_c  = 2 \Sigma_{bac} - 2 \Sigma_{cab} , \quad 
t_c = \frac{2}{2-n}\Sigma_c \nn 
\eeq
Therefore, 
\beq
 t_{abc} =  2\Sigma_{bac} -  2\Sigma_{cab} - \frac{2}{n-2} (\eta_{ab}\Sigma_c - \eta_{ac}\Sigma_b) 
 \nn
\eeq
Using the constraints (\ref{ce}), 
\begin{align}
\ge \Sigma_{a}{}^{bc} &\approx {-4\pi G}\, e_{ [\alpha }^b e^c_{ \beta  ]} p^\alpha_{e^a_\beta}  \nn \\
 \ge \Sigma^c 
  %& =\delta^a_b \Sigma_{a}{}^{bc} 
  &\approx {-4\pi G}\, e^a_{[\alpha} e^c_{\beta ]} p^\alpha_{e^a_\beta} \nn
\end{align}
Then 
\begin{align}  \label{fabc}
 \ge t_{abc} \approx {8\pi G}\Big(  & e_{a [\alpha } e_{b| \beta  ]} p^\alpha_{e^c_\beta} -   e_{a [\alpha } e_{c| \beta  ]} p^\alpha_{e^b_\beta}
 + \frac{1}{n-2}\big( \eta_{a b} e^d_{[\alpha} e_{c|\beta ]} p^\alpha_{e^d_{\beta}} 
  - \eta_{a c} e^d_{[\alpha} e_{b |\beta ]} p^\alpha_{e^d_{\beta}} \big) \Big)
  %+  \frac{2 }{n-2} \eta_{a [b} e^d_{[\alpha} e_{c]\beta ]} p^\alpha_{e^d_{\beta}} 
% \nn \\  &  + \frac{1}{n-2}\big( \eta_{a b} e^d_{[\alpha} e_{c|\beta ]} p^\alpha_{e^d_{\beta}} 
%  - \eta_{a c} e^d_{[\alpha} e_{b |\beta ]} p^\alpha_{e^d_{\beta}} \big) \Big)
\end{align}

The DW Hamiltonian density (\ref{dwhprim}) can be expressed now as a function of the variables of the reduced polymomentum phase space as follows 
\beq
\label{hdwr}
 \ge H  \approx \frac14 p^\alpha_{e^a_\beta} e_\alpha^b e_\beta^c t^a{}_{bc} % \nn \\
%& \approx 
% {2\pi G} \ge^{-1} p^\alpha_{e^a_\beta} e_\alpha^b e_\beta^c \left( -   e_{ [\alpha }^a e_{c|\beta]} 
% p^\alpha_{e^b_\beta}
%+  e_{ [\alpha }^a e_{b |\beta  ]} p^\alpha_{e^c_\beta}  
%+  \frac{2}{n-2} \eta^a_{b} e^d_{[\alpha} e_{c |\beta ]} p^\alpha_{e^d_\beta}\right)
%\nn \\
 \approx {4\pi G} \ge^{-1} p^\alpha_{e^a_\beta} e_{[\alpha}^b e_{\beta]}^c \left(
  e_{ [\alpha }^a e_{b |\beta  ]} p^\alpha_{e^c_\beta}  
+  \frac{1}{n-2} \eta^a_{b} e^d_{[\alpha} e_{c |\beta ]} p^\alpha_{e^d_\beta} 
\right) 
\eeq

%\subsection{The field equations in DW Hamiltonian form}  

\section{Covariant DW Hamilton-Jacobi equation}  

In general space-times the covariant DW Hamilton-Jacobi equation is formulated in terms of the 
eikonal densities $S^\mu$ and the DW Hamiltonian function in (\ref{dwhjeq}) is replaced by the 
DW Hamiltonian density $\ge H$. The Dirac bracket (\ref{pee}) shows that the polysymplectic structure on the reduced  polymomentum phase space is the standard one 
\beq
\Omega^{red} = d e\we dp^\alpha_e \we \upsilon_\alpha
\eeq 
or, respectively, the related $k-$symplectic  structure used in \cite{dlhj2}  is given by the family of forms $ d e\we dp^\alpha_e $. 

This observation allows us to use the geomteric theory of HJ equations developed in \cite{dlhj1,dlhj2,dlhj3} to write down the covariant DW Hamilton-Jacobi equation in the form 
\beq \label{tpembed}
\der_\mu S^\mu + \ge H \left(e, p^\alpha_e= \frac{\der S^\alpha}{\der e} , x \right) = 0
\eeq
in which the expression (\ref{hdwr}) of $H$ on the reduced polymomentum phase space in used. 
Then the theorems proven in \cite{dlhj1,dlhj2,dlhj3}  guarantee that the embedding condition 
\beq
p^\alpha_e (e(x), \der_\alpha e(x), x^\alpha) := \frac{\der S^\alpha}{\der e} 
\eeq
describes the classical solutions of the teleparallel equivalent of the Einstein equations which 
are derived from the variational  principle based on (\ref{lagrtp}). 
% Note that both $S^\alpha$ and $p^\alpha_e$ are densities. 
Due to the constraint (10) and the fact that on classical solutions $t_{abc} = T_{abc}(x)$ 
the embedding condition has the explicit form
\beq \label{tpembed}
\frac{1}{2\pi G} \ge \Sigma(T)_a{}^{cb} e^\alpha_{b}e^\beta_{c} 
= \frac{\der S^{\alpha}}{\der e^a_{\beta}} - \frac{\der S^{\beta}}{\der e^a_{\alpha}} ~.
\eeq 
which is valid along the solutions $e = e(x)$. 

\section{Conclusion} 
Using the Kanatchikov's algorithm for the treatment of constraints within the covariant De Donder-Weyl Hamiltonian formulation we have constracted the DW Hamiltonian density of the Palatini formulation 
of the teleparallel equivalent of General Relativity on the reduced polymomentum phase space 
and formulated the covariant DW Hamilton-Jacobi equation which is defined on the configuration space 
of vielbein variables and space-time variable. The problems to be considered in future work is 
a generalization of our consideration to non-Einsteinian teleparallel gravity theories and  
the study of the relation of our covariant Hamilton-Jacobi formulation with the standard Hamilton-Jacobi formulation based on 3+1 decomposition which is not yet formulated within the canonical Hamiltonian 
formalism for teleparallel gravities. It is also interesting to investigate if the DW Hamilton-Jacobi equation we have derived can be obtained in the quasiclassical limit of the precanonical quantization 
of TEGR described in [59,60].

 \footnotesize 
%\section*{Acknowledgments}
%To thank St Andrews U, if there is a space left!


\vspace*{-5pt} 



\begin{thebibliography}{99}

\bibitem{dedonder} Th. de Donder, Th\'eorie invariantive du calcul des variations (Gauthier‐Villars, Paris, 1930), %% or 1935 nouv. éd.

\bibitem{horava} P. Ho\v{r}ava, On a covariant Hamilton-Jacobi framework for the Einstein-Maxwell theory,
Class. Quant. Grav. 8 (1991) 2069.  %-2084 • DOI: 10.1088/0264-9381/8/11/016

\bibitem{weyl} H. Weyl, Geodesic fields in the calculus of variation for multiple integrals,
Ann. Math. 36 (1935) 607.  %(1935).

\bibitem{volterra} V. Volterra, Sopra una estensione della teoria Jacobi–Hamilton del calcolo delle varizioni, Atti Accad. Naz. Lincei Rend. (Sec. IV) VI  (1890) 127. 
%Rend. Cont. Acad. Lincei, ser. IV, vol. VI  (1890) 127. %–138.

 \bibitem{carath} C. Carath\'eodory, 
 \"Uber die Variationsrechnung bei mehrfachen Integralen, 
 Acta Szeged 4 (1929) 193. %–216 (1929).
 
 \bibitem{prange} G. Prange, Die Hamilton-Jacobische Theorie f\"ur Doppelintegrale, Dissertation
 G\"ottingen (1915). 
 
 \bibitem{boerner} H. Boerner, %munich 
 \"Uber die Extremalen und geodätischen Felder in der Variationsrechnung der mehrfachen Integrale, 
  Math. Ann. 112 (1936) 187. %–220 (1936).
 
%J. Ernest Wilkins, Jr. (1944). Multiple Integral Problems in Parametric Form in the Calculus of %Variations. Annals of Mathematics (Second Series), 45(2), 312–334. doi:10.2307/1969268  
 
 \bibitem{rund} H. Rund, The Hamilton-Jacobi theory in the calculus of variations, 
 D. van Nostrand, London and New York, 1966.
 
%\bibitem{hjbook} M. Giaquinta and S. Hildebrandt, Calculus of Variations, vols. I and II, Springer 2013. 

 
\bibitem{kastrup} H. Kastrup, Canonical theories of Lagrangian dynamical systems in physics, 
Phys. Rep. 101 (1983) 1. %Issues 1–2, December 1983, Pages 1-167
%


\bibitem{nambu} Y. Nambu, Hamilton-Jacobi formalism for strings, 
 %Part of Broken symmetry: Selected papers of Y. Nambu
Phys. Lett. B 92 (1980) 327. 

%String mechanics based on 2‐forms
%J. Math. Phys. 23, 388 (1982); https://doi.org/10.1063/1.525381
%P. Mitra
%Hamiltonian Constraint Formulation of Classical Field Theories
%V. Zatloukal 
%Advances in Applied Clifford Algebras volume 27, pages829–851 (2017)

\bibitem{kastruprinke} H. Kastrup and M. Rinke,  
Hamilton-Jacobi theories for strings, 
Phys. Lett.  B105 (1981) 191. 

\bibitem{rinke}
M. Rinke, 
The relation between relativistic strings and Maxwell fields of rank 2, 
Comm. Math. Phys.   73 (1980) 265. %-71 

\bibitem{ikan-string} I.V. Kanatchikov, 
Weyl Spinor Field from the Hamilton-Jacobi Formulation of Null Strings,
 Europhys. Lett. 12 (1990) 577. 
 
\bibitem{dlhj1} M. de Le\'on,  J. C. Marrero, D. Martin De Diego,
A Geometric Hamilton-Jacobi Theory for Classical Field Theories, 
%In this paper we extend the geometric formalism of the Hamilton-Jacobi theory for hamiltonian mechanics to the case of classical field theories in the framework of multisymplectic geometry and Ehresmann connections.
	arXiv:0801.1181 [math-ph]

\bibitem{dlhj2}
M. de Le\'on, D. Martin De Diego, J. C. Marrero, M. Salgado, S. Vilari\~no, 
Hamilton-Jacobi Theory in k-Symplectic Field Theories,
Int. J. Geom. Meth. Mod. Phys. 7 (2010) 1491, %-1507
e-Print: 1005.1496 [math-ph]
%DOI: 10.1142/S0219887810004919

\bibitem{dlhj3} 
M. de Le\'on, P. D. Prieto-Mart\'inez, N. Rom\'an-Roy, S.Vilari\~no, 
J. Math. Phys. 58 (2017) 092901.  
arXiv:1504.02020. % DOI: 10.1063/1.5004260

\bibitem{rund89} H. Rund, Legendre transformations and Cartan forms in the calculus of variations of multiple integrals: Part I and II, Quaest. Mathem. 12 (1989) 205, 315. %-229
%Part I: Legendre Transformations
%\bibitem{rund2}
%Part II: Canonical Distributions and Cartan Forms, 
%LEGENDRE TRANSFORMATIONS AND CARTAN FORMS IN THE CALCULUS OF VARIATIONS OF MULTIPLE INTEGRALS: Part I: Legendre Transformations
%H Rund - Quaestiones Mathematicae, 1989 - Taylor $\&$ Francis
%LEGENDRE TRANSFORMATIONS AND CARTAN FORMS IN THE CALCULUS OF VARIATIONS OF MULTIPLE INTEGRALS: Part 11: Canonical Distributions and …
%H Rund - Quaestiones Mathematicae, 1989 - Taylor $\&$ Francis

%\bibitem{bertin-hj}
%M.C. Bertin, B.M. Pimentel, P.J. Pompeia,
%Hamilton-Jacobi approach for first order actions and theories with higher derivatives,
%Ann. Phys. 323 (2008) 527-547, arXiv: hep-th/0701262. %• DOI: 10.1016/j.aop.2007.11.003

\bibitem{bruno-hj} D. Bruno, 
Constructing a class of solutions for the Hamilton-Jacobi equation in field theory,
J. Math. Phys. 48 (2007) 112902, arXiv:0709.1930 % DOI: 10.1063/1.2804076

\bibitem{rr21}  N. Rom\'an-Roy,  An overview of the Hamilton--Jacobi theory: the classical and geometrical approaches and some extensions and applications, 
	Mathematics 9 (2021) 85, 
%(Special Issue: "New Trends in Hamilton-Jacobi Theory: Conservative and Dissipative Dynamics")
%DOI:	10.3390/math9010085
arXiv:2101.03830 [math-ph]. 

%\bibitem{vonrieth} J. von Rieth,
%The Hamilton–Jacobi theory of De Donder and Weyl applied to some relativistic field theories,
%J. Math. Phys. 25 (1984) 1102. % https://doi.org/10.1063/1.526253




\bibitem{weiss} P. Weiss, 
On the Hamilton-Jacobi theory and quantization of a dynamical continuum, 
 Proc. R. Soc. Lond. A169  (1938) 102.  %–119
%http://doi.org/10.1098/rspa.1938.0197
%On the Hamilton-Jacobi Theory and Quantization of Generalized Electrodynamics (pp. 119-133)
%P. Weiss
%https://www.jstor.org/stable/97248
%Weiss A156 (1936) 192. -- schr functiopnal
%ON WEISS'S THEORY OF FIELDS
%T. S. CHANG Acta Physica Sinica, 1949, 7(4): 59-71. doi: 10.7498/aps.7.59-3
  %The Parisi formula is a Hamilton–Jacobi equation in Wasserstein space
  %Jean-Christophe Mourrat  2021
  %Canadian Journal of Mathematics , First View , pp. 1 - 23
  %DOI: https://doi.org/10.4153/S0008414X21000031


\bibitem{hj-thesis} J.R. Pessoa de Ara\'ujo, %João Ricardo Pessoa de Araújo 
Formalismo de Hamilton-Jacobi Aplicado a Teorias de Campos Topol\'ogicas,  
PhD Thesis, Universidade Federal da Bahia, Salvador, Brazil (2010). 
%UNIVERSIDADE FEDERAL DA BAHIA
%https://ppgfis.ufba.br/sites/ppgfis.ufba.br/files/dissertacao_joao_ricardo.pdf
\bibitem{ik-pla} I. V. Kanatchikov, 
Precanonical Quantization and the Schroedinger Wave Functional, 
Phys. Lett. A283 (2001) 25, %-36 
	arXiv:hep-th/0012084.
%DOI:	10.1016/S0375-9601(01)00225-0
\bibitem{nikolic}  H. Nikolic, 
Covariant canonical quantization of fields and Bohmian mechanics, 
 Eur. Phys. J. C 42 (2005) 365, %-374
arXiv:hep-th/0407228 [hep-th]. 
%Quantum determinism from quantum general covariance, 
%Int. J. Mod. Phys. D 15 (2006) 2171, %-2176
%arXiv:hep-th/0601027. 

\bibitem{riahi} 
N. Riahi and M. E. Pietrzyk, 
On the Relation Between the Canonical Hamilton–Jacobi Equation and the De Donder–Weyl Hamilton–Jacobi Formulation in General Relativity, 
Acta Phys. Polon. Supp. 13 (2020) 213.

\bibitem{rovelli} C. Rovelli, 
Dynamics without time for quantum gravity: Covariant Hamiltonian formalism and Hamilton-Jacobi equation on the space $G$, Lect. Notes Phys. 633 (2003) 36,  %-62
arXiv:gr-qc/0207043. 

\bibitem{conrady} F. Conrady and C. Rovelli, 
Generalized Schroedinger equation in Euclidean field theory,
Int. J. Mod. Phys. A19 (2004) 4037,  %-4068
 	arXiv:hep-th/0310246. 
 	
 \bibitem{dopli} L. Doplicher, 
 Generalized Tomonaga-Schwinger equation from the Hadamard formula, 
 Phys.Rev. D 70 (2004) 064037, 
 arXiv:gr-qc/0405006.  
 
 

 
 %the programme of precanonical quantization which started in \cite{ik1,ik2,ik3,ik5e}
\bibitem{ik1} I. Kanatchikov, 
From the Poincar\'e-Cartan form to a Gerstenhaber algebra of Poisson brackets in field theory, 
arXiv:hep-th/9511039.  
\bibitem{ik2} I. Kanatchikov, 
Towards the Born-Weyl Quantization of Fields
arXiv:quant-ph/9712058.  
\bibitem{ik3} I. V. Kanatchikov, 
DeDonder-Weyl theory and a hypercomplex extension of quantum mechanics to field theory, 
Rept. Math. Phys. 43 (1999) 157,  %-170
doi:10.1016/S0034-4877(99)80024-X,
arXiv:hep-th/9810165.  
\bibitem{ik4} I. V. Kanatchikov, 
On Quantization of Field Theories in Polymomentum Variables, 
AIP Conf. Proc. 453 (1998) 356, %-367,1998
doi:10.1063/1.57105,
arXiv:hep-th/9811016.
\bibitem{ik5} I. V. Kanatchikov, 
Geometric (pre)quantization in the polysymplectic approach to field theory, 
arXiv:hep-th/0112263. 

\bibitem{ik5e} I. V. Kanatchikov, Ehrenfest theorem in precanonical quantization, 
J. Geom. Symmetry Phys. 37 (2015) 43, %-66
doi:10.7546/jgsp-37-2015-43-66, 
arXiv:1501.00480.  


%%earlier discovery of the bi-graded (Gerstenhaber) analogue of the Poisson bracket in the 
%simplest De Donder-Weyl covariant Hamiltonian formalism\cite{ikbr1,ikbr2,ikbr3,khbr1,khbr2,khbr3}.

\bibitem{ikbr1} I. V. Kanatchikov, 
On the Canonical Structure of De Donder-Weyl Covariant Hamiltonian 
Formulation of Field Theory 1. Graded Poisson Brackets and Equations of Motion, 
 arXiv:hep-th/9312162. 
 
\bibitem{ikbr2}  
I. Kanatchikov,   
Canonical Structure of Classical Field Theory in the Polymomentum Phase Space, 
 Rep. Math. Phys.  41 (1998) 49,  
 arXiv:hep-th/9709229. 
 
\bibitem{ikbr3} 
I. Kanatchikov, 	
  On Field Theoretic Generalizations of a Poisson Algebra,  
 Rep. Math. Phys. { 40} (1997) 225,  
 {  arXiv:hep-th/9710069}.

%\bibitem{ikbr3} ... Loday

\bibitem{khbr1} F. H\'elein, J. Kouneiher,
Finite dimensional Hamiltonian formalism for gauge and quantum field theories,
J. Math. Phys. 43  (2002) 2306. %-2347
%DOI:10.1063/1.1467710

\bibitem{khbr2} F. H\'elein, J. Kouneiher, 
Covariant Hamiltonian formalism for the calculus of variations with several variables, 
arXiv:math-ph/0211046

\bibitem{khbr3} F. H\'elein, J. Kouneiher, 
Covariant Hamiltonian formalism for the calculus of variations with several variables: 
Lepage-Dedecker versus de Donder-Weyl, 
Adv. Theor. Math. Phys. 8 (2004) 565.  %565-601

%arXiv:1406.3641  [pdf, ps, other]  math-ph math.DG
%Multisymplectic formulation of Yang--Mills equations and Ehresmann connections
%Authors: Frédéric Hélein




%to gauge fields \cite{ik5e,iky1,iky2,iky3}

\bibitem{iky1} I. V. Kanatchikov, 
Precanonical quantization of Yang-Mills fields and the functional Schroedinger representation, 
Rept. Math. Phys. 53 (2004) 181, %-193
%doi:10.1016/S0034-4877(04)90011-0, 
arXiv:hep-th/0301001. 

\bibitem{iky2} I. V. Kanatchikov, 
On the spectrum of DW Hamiltonian of quantum SU(2) gauge field
Int. J. Geom. Methods Mod. Phys. 14 (2017) 1750123, 
%doi:10.1142/S0219887817501237, 
arXiv:1706.01766. 

\bibitem{iky3} I. V. Kanatchikov, 
Schr\"odinger wave functional in quantum Yang-Mills theory from precanonical quantization, 
Rep. Math. Phys. 82 (2018) 373, 
%doi:10.1016/S0034-4877(19)30008-4, 
arXiv:1805.05279. 



%metric formulation  \cite{ikm1,ikm2,ikm3,ikm4} 

\bibitem{ikm1} I. V. Kanatchikov, 
From the DeDonder-Weyl Hamiltonian formalism to quantization of Gravity, 
	arXiv:gr-qc/9810076

\bibitem{ikm2} I. V. Kanatchikov, Quantization of Gravity: Yet Another Way, 
arXiv:gr-qc/9912094  


\bibitem{ikm3} I. V. Kanatchikov, 
Precanonical Perspective in Quantum Gravity
Nucl. Phys. Proc. Suppl. 88 (2000) 326, %-330
 %doi:10.1016/S0920-5632(00)00795-7, 
arXiv:gr-qc/0004066.  
 
\bibitem{ikm4} I. V. Kanatchikov, 
Precanonical Quantum Gravity: quantization without the space-time decomposition, 
Int. J. Theor. Phys. 40 (2001) 1121,   %-1149
arXiv:gr-qc/0012074. 
%doi 10.1023/A:1017557603606

%Palatini vielbein formulation \cite{ikv1,ikv2,ikv3,ikv4,ikv5}. 

\bibitem{ikv1} 
  %\bibitem{pqg-vielbein}   
I. V. Kanatchikov, 
{ On Precanonical Quantization of \ Gravity in Spin Connection Variables}, 
{\rm AIP Conf. Proc. } { 1514} (2012) 73, % (2012), % 73-76,  
{\rm arXiv:1212.6963} [gr-qc].
\bibitem{ikv2} 
I. V. Kanatchikov, 
{  De Donder-Weyl Hamiltonian Formulation and Precanonical Quantization 
of \ Vielbein Gravity}, 
{\em J. Phys. Conf. Ser.} { 442} (2013) 012041, 
{\rm arXiv:1302.2610} [gr-qc]. 
\bibitem{ikv3} 
I. V. Kanatchikov, {\rm  On Precanonical Quantization of \ Gravity}, 
 {\rm Nonlin. Phenom. Complex Sys. (NPCS)} 
  {\rm 17} (2014) 372, %(2014). %372-376, 
 arXiv:1407.3101 [gr-qc]. 


\bibitem{ikv4} I. V. Kanatchikov, Ehrenfest Theorem in Precanonical Quantization of Fields and Gravity,
arXiv:1602.01083  

\bibitem{ikv5} I. V. Kanatchikov, On the ``spin connection foam" picture of quantum gravity from precanonical quantization, 
arXiv:1512.09137  


%Schr\"odinger representation has been established in \cite{ik-pla,iks1,iks2,iksc1,iksc2,iksc3} 
%both for quantum scalar fields in flat and curved space-times and Yang-Mills gauge theory. 

\bibitem{iks1} I. V. Kanatchikov, Precanonical quantization and the Schrödinger wave functional revisited,
Adv. Theor. Math. Phys. 18 (2014) 1249, %-1265
arXiv:1112.5801  .

\bibitem{iks2} I. V. Kanatchikov, On the precanonical structure of the Schr\"odinger wave functional,
Adv. Theor. Math. Phys. 20 (2016) 1377, %-1396
arXiv:1312.4518. 

%\bibitem{iks3} I. V. Kanatchikov, 



\bibitem{iksc1}
I. V. Kanatchikov, Schr\"odinger Functional of a Quantum Scalar Field in Static Space-Times from Precanonical Quantization, 
Int. J. Geom. Meth. Mod. Phys. 16 (2019) 1950017, 
arXiv:1810.09968.   

\bibitem{iksc2}
I. V. Kanatchikov, Precanonical structure of the Schr\"odinger wave functional in curved space-time,
Symmetry 11 (2019) 1413, %https://doi.org/10.3390/sym11111413
arXiv:1812.11264.

\bibitem{iksc3} I. V. Kanatchikov, 
On the precanonical structure of the Schr\"odinger wave functional in curved space-time, 
Acta Phys. Polon. B Proc. Suppl. 13 (2020) 313, 
arXiv:1912.07401  


\bibitem{guiding} M. Derakhshani, M. K.-H. Kiessling and A. S. Tahvildar-Zadeh, 
Covariant Guiding Laws for Fields, 
arXiv:2110.09683. 


\bibitem{peres2} A. Peres, On Cauchy’s problem in general relativity - II, 
Il Nuovo Cimento (1955-1965) 26 (1962) 53. %%pages53–62 (1962). 

\bibitem{ik-tpq} I. V. Kanatchikov, Is quantum TEGR equivalent to quantum GR?, 
a talk at "Alternative Gravities and Fundamental Cosmology - ALTECOSMOFUN'21", 
Szeczecin 2021. 

\bibitem{ik-mg21} I. V. Kanatchikov, 
Precanonical quantization of Teleparallel Equivalent of General Relativity
a talk at the Sixteenth Marcel Grossmann Meeting, Rome 2021. 
  

\bibitem{gerlach69} U. H. Gerlach, 
Derivation of the Ten Einstein Field Equations from the Semiclassical
Approximation to Quantum Geometrodynamics,
Phys. Rev. 177 (1969) 1929. 


\bibitem{maluf2} 
J. W. Maluf, The teleparallel equivalent of general relativity. 
{ Ann. der Phys.} { 525}  339 (2013) %-357 
{arXiv:gr-qc/1303.3897}.
%doi:10.1002/andp.201200272.

\bibitem{jhk} J. B. Jimenez, L. Heisenberg and T. S. Koivisto,  
 Teleparallel Palatini theories, 
 	arXiv:1803.10185 [gr-qc]. 


%{\em J. Cosmol. Astropart. Phys.} 2018 (08)???
%DOI:10.1088/1475-7516/2018/08/039

%\bibitem{hoh} M. Hohmann, 
%Variational Principles in Teleparallel Gravity Theories, 
%	Universe 7 (2021) 114
%DOI:	10.3390/universe7050114
%arXiv:2104.00536.
	

\bibitem{maluf} 
J. W. Maluf and J.F. da Rocha-Neto, 
Hamiltonian formulation of general relativity in the teleparallel geometry,
{\em Phys. Rev.} {\bf D64} (2001) 084014, 
 	arXiv:gr-qc/0002059 %it has a different title below!
%J. W. Maluf  and J. F. da Rocha-Neto
%Hamiltonian formulation of the teleparallel equivalent of general relativity without gauge fixing
  %Palatini eq 5!
%gr-qc/0002059 [gr-qc]

 \bibitem{butterfield} 
J. Butterfield, On Hamilton-Jacobi Theory as a Classical Root of Quantum Theory,  
in  "Quo Vadis Quantum Mechanics? Possible Developments in Quantum Theory in the 21st Century," 
Elitzur, A.C., Dolev, S., and Kolenda, N. (editors),
New York: Springer (2004) 239. %2004 p.239-273 
%On Hamilton-Jacobi theory: its geometry and relation to pilot-wave theory
%Jeremy Butterfield	arXiv:quant-ph/0210140
	
	
\bibitem{dirac} P. A.M. Dirac, 
	Lectures on Quantum Mechanics,  Dover Publications (1964). 	
	
\bibitem{ik-dirac} 
I.V. Kanatchikov,
{  On a Generalization of the Dirac Bracket 
in the De Donder-Weyl Hamiltonian Formalism}, 
In: {Differential Geometry and its Applications, } 
 %Proc. 10th Int. Conf. on Diff. Geom. \& Appl., Olomouc, August 2007, 
 Kowalski O., Krupka D., Krupkov\'a O. and Slov\'ak J. (Eds), 
 World Scientific, Singapore 2008, 615,  %pp 615-625, 
 arXiv:0807.3127. 
	
	
\bibitem{mx1} 	J. Berra-Montiel, A. Molgado, D. Serrano-Blanco, 
	De Donder-Weyl Hamiltonian formalism of MacDowell-Mansouri gravity, 
Class. Quantum Grav. 34 (2017) 235002, 	
arXiv:1703.09755.  
%doi 10.1088/1361-6382/aa924a

\bibitem{mx2} 	J. Berra-Montiel, A. Molgado, A.  Rodriguez-Lopez,  
Polysymplectic formulation for BF gravity with Immirzi parameter, 
Class. Quantum Grav. 36 (2019) 115003, 
arXiv:1901.11532. 
\bibitem{mx3} 	J. Berra-Montiel, A. Molgado, A.  Rodriguez-Lopez,  
Class. Quantum Grav. 38 (2012) 135012, 
arXiv:2101.08960  
%doi 10.1088/1361-6382/abf711


\end{thebibliography}
\end{document}